# Nanoscale phase separation in the iron chalcogenide superconductor $K_{0.8}Fe_{1.6}Se_2$ as seen via scanning nanofocused x-ray diffraction

A. Ricci[1], N. Poccia[1], G. Campi[2], B. Joseph[1,3], G. Arrighetti[4], L. Barba[4], M. Reynolds[5], M. Burghammer[5], H. Takeya[3], Y. Mizuguchi[3], Y. Takano[3], M. Colapietro[2], N. L. Saini[1], A. Bianconi[1]

[1]Department of Physics, Sapienza University of Rome, P. le A. Moro 2, 00185 Rome, Italy

[2]Institute of Crystallography, CNR, via Salaria Km 29.300, Monterotondo Stazione, Roma, I-00016, Italy

[3]National Institute of Material Science, 1-2-3 Sengen, Tsukuba, 3050047 Japan.

[4]Elettra Sincrotrone Trieste. Strada Statale 14 - km 163,5, AREA Science Park, 34149 Basovizza, Trieste, Italy

[5]European Synchrotron Radiation Facility, B. P. 220, F-38043 Grenoble Cedex, France

*E-mail: antonio.bianconi@roma1.infn.it

Advanced synchrotron radiation focusing down to a size of 300 *nm* has been used to visualize nanoscale phase separation in the $K_{0.8}Fe_{1.6}Se_2$ superconducting system using scanning nanofocus single-crystal X-ray diffraction. The results show an intrinsic phase separation in $K_{0.8}Fe_{1.6}Se_2$ single crystals at T< 520 K, revealing coexistence of i) a magnetic phase characterized by an expanded lattice with superstructures due to Fe vacancy ordering and ii) a non-magnetic phase with an in-plane compressed lattice. The spatial distribution of the two phases at 300 K shows a frustrated or arrested nature of the phase separation. The space-resolved imaging of the phase separation permitted us to provide a direct evidence of nanophase domains smaller than 300 *nm* and different micrometer-sized regions with percolating magnetic or nonmagnetic domains forming a multiscale complex network of the two phases.

Superconductivity, with transition temperature, $T_c$, above 30 K, has been reported in the new family of high temperature superconductors (HTSs) $A_xFe_{2-y}Se_2$ [1-6]. These systems are made of iron chalcogenide FeSe molecular layers, intercalated by A=K,Cs,Rb,Tl, (Tl,Rb), (Tl,K) spacer layers, providing the more recent practical realization of metal heterostructures at atomic limit, as cuprates and pnictides high-temperature superconductors [7]. These $A_xFe_{2-y}Se_2$ chalcogenide superconductors show both high-temperature superconductivity and magnetism [8-14]. In these compounds, one may tune the interplay of superconductivity and magnetism by changing the Fe-vacancy order and the superlattice misfit strain [15]. The open question to be answered is whether there is a co-existence of magnetism and superconductivity in the same spatial region or these phenomena occur in different spatial regions dictated by the phase separation. Opposite conclusions have been drawn by different groups using different experimental methods, some favouring the coexistence [7-11], while others the phase separation [12-14]. The phase separation appears on multiple scales from the micrometer scale to nanoscale in cuprates [16-19], diborides [20] and pnictides [21]. Several theories [22-30] have described the complex phase separation as an intrinsic feature of all known HTSs.

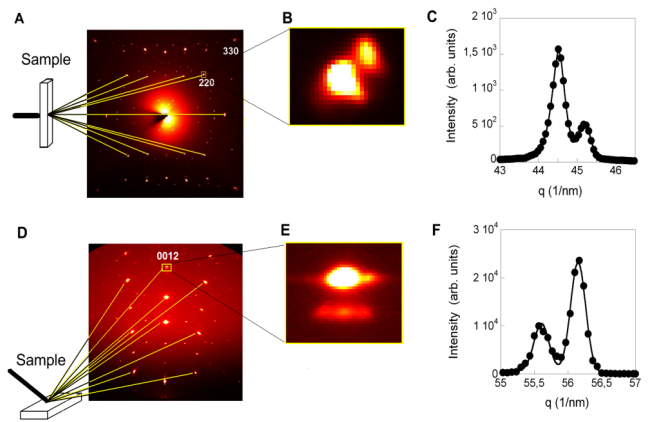

**Figure 1:** (Color online) (A) X-ray diffraction setup in transmission mode at the XRD1 beamline at the ELETTRA synchrotron radiation facility providing the CCD image of the x-rays reflections on the $K_{0.8}Fe_{1.6}Se_2$ obtained at room temperature. A zoom over the (220) peak is shown in (B), revealing coexisting phases in $K_{0.8}Fe_{1.6}Se_2$ evidenced by the splitting of the (220), one with a smaller in-plane lattice parameters (called the compressed phase) and one with the longer lattice parameter (called the expanded phase). (C) Diffraction profile of the (220) reflection is shown. X-ray diffraction image obtained in reflection mode is shown in (D). A zoom over the (0012) reflections is also shown (E). Peak profile of the (0012) reflection in the c direction is shown, revealing a clear phase separation in the c axis (F).

X-ray diffraction (XRD), the most direct probe of phase separation, has revealed a phase separation in undoped magnetic phase and nonmagnetic phase in an underdoped $La_2CuO_{4+y}$ system [17]. The phase separation in an overdoped $La_2CuO_{4+y}$ system has been observed recently by scanning micro-XRD [19] to simultaneously probe both real and reciprocal spaces. Here we report evidence of frustrated phase separation in $K_{0.8}Fe_{1.6}Se_2$ with domains of magnetic phase, characterized by expanded lattice with $\sqrt{5}\times\sqrt{5}\times1$ ordered superstructure of iron vacancies, and domains of in-plane compressed nonmagnetic phase. Using conventional high-energy XRD and advances in focusing x-ray synchrotron radiation we have been able to probe both the average bulk structure and to map of the phase separation on a nanometre scale in the same sample.

The experimental set-up used for the single-crystal diffraction studies in the transmission mode (probing the average bulk structure) and the reflection mode (nanofocus beam scanning to map the phase separation) is shown in Fig 1. The single crystals of $K_{0.8}Fe_{1.6}Se_2$ were grown by melting a precursor of FeSe and K (in nominal composition of 2:0.8) placed in an alumina crucible that was sealed into an arc-welded stainless-steel tube. The sample went through a heating treatment at 1030 ºC for 2 h followed by a slow cooling down to 750 ºC with a rate of 6 ºC/h. We obtained platelike and dark-shining single crystals. The details on the crystal growth can be found in ref. [6]. The obtained crystals were characterized for their phase purity using in-house XRD measurements. The actual atomic composition of the crystals was determined by energy-dispersive X-ray spectroscopy (EDX) analyzing the X-ray fluorescence lines of K, Fe and Se excited by the electron beam. The composition has been found to be K:Fe:Se = 0.8:1.6:2 using an average of four points. Temperature-dependent magnetization and resistivity measurements were performed using a superconduction quantum interference device (SQUID) magnetometer and a four-probe system respectively. The single crystals show a sharp superconducting transition temperature ~32 K. For the synchrotron radiation XRD measurements, reported in this Rapid Communication, we used a small piece of a large platelike single crystal with a $T_c$ = 31.8 K, that was well characterized for its phase purity and composition. The single-crystal diffraction profiles were found to be very sharp, ascertaining high quality of the crystal used in the present study.

We first present results on the phase separation observed in the temperature-dependent (80-600 K) single-crystal XRD (in transmission mode), probing a $105 \mu m$ depth bulk sample, i.e., a volume of $200 \times 200 \times 105 \mu m^3$. Fig 2 shows the results of the temperature dependent single-crystal XRD measurements, conducted at the XRD1 beamline of the ELETTRA synchrotron radiation facility, Trieste, using photons of energy 20 KeV.

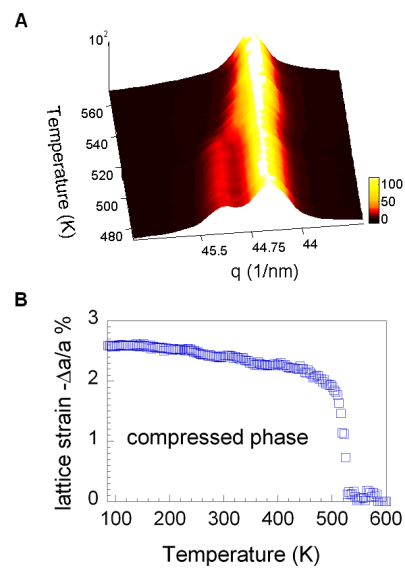

**Figure 2:** (Color online) (A) A three-dimensional color plot of the temperature dependence of the (220) reflection measured on the $K_{0.8}Fe_{1.6}Se_2$ single crystal. A clear phase separation in a compressed phase (shorter a,b parameters) phase and an expanded (longer a,b lattice parameters) phase appears below 520 K. (B) Temperature dependence of in-plane difference of the tetragonal a,b lattice parameters between the compressed and the expanded phase recorded with a cooling rate of 0.1K/min.

The crystal symmetry at 600 K is found to be of tetragonal (space group I4/mmm, $ThCr_2Si_2$-type tetragonal unit-cell) with lattice parameters a=b=0.401 nm and c=1.384 nm. Below 580 K, superlattice peaks start to appear indicating Fe vacancy ordering. This vacancy-ordered structure, called the A phase, and described by several groups [11-15], has been associated with the magnetic order [8-10].

The system undergoes an intrinsic phase separation below 520 K with the appearance of a new phase with slightly compressed basal lattice parameters (a, b). The evolution of lattice parameters are reported in Ref. 31 and Fig. 2 shows the splitting of the (220) reflection at variable temperature. The appearing phase has 2.5% shorter in-plane lattice parameters and ~1% expanded c-axis, called the in-plane compressed phase (B phase). The probability of the phase B turns out to be ~20-30% averaged over a volume of $4.2 \times 10^{15} \, nm^3$.

To study the nanoscopic spatial distribution of the intrinsic phase separation, we have obtained sample imaging at room temperature after a slow cooling rate of 0.1K/min from 530K. We have measured the XRD in the reflection mode using a 300 × 300 nm² beam size with a penetration depth of reflected x-rays of ~$11 \mu m$.

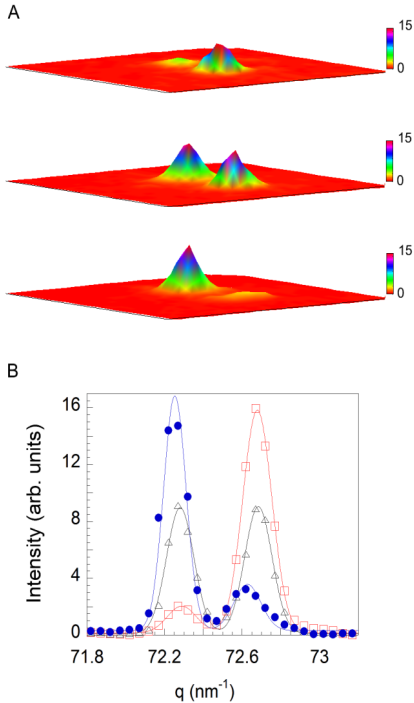

**Figure 3:** (Color online) (A) The reciprocal three-dimensional view of the x-ray reflections around (0016) of the $K_{0.8}Fe_{1.6}Se_2$ measured at the ESRF using a 300-nm focused beam. Reflections from three different regions of the sample are shown, revealing the variable strength of the expanded and compressed phases at different spatial locations. (B) The corresponding two-dimensional profiles along the c* direction are shown. The diffraction coherence length estimated from the full width at half maximum (FWHM) of the reflection profiles is found to be $\xi^l = 6 \pm 1 nm$.

Thus the volume of the probed surface for a single diffraction pattern is ~$9.9 \times 10^8$ nm³, i.e., ~$2.3 \times 10^7$ times smaller than the volume probed by the transmission XRD. Fig. 3 displays the (0016) reflections at three different positions of area (300 × 300 nm²) each on the single-crystal $K_{0.8}Fe_{1.6}Se_2$ surface at 300 K. Within the same $9.9 \times 10^8$ nm³ crystal surface-volume illuminated by the x-ray beam, we probe a different phase separation inside the illuminated volume given by the ratio between the (0016) reflection intensities due to the A and B phases, showing large changes from site to site.

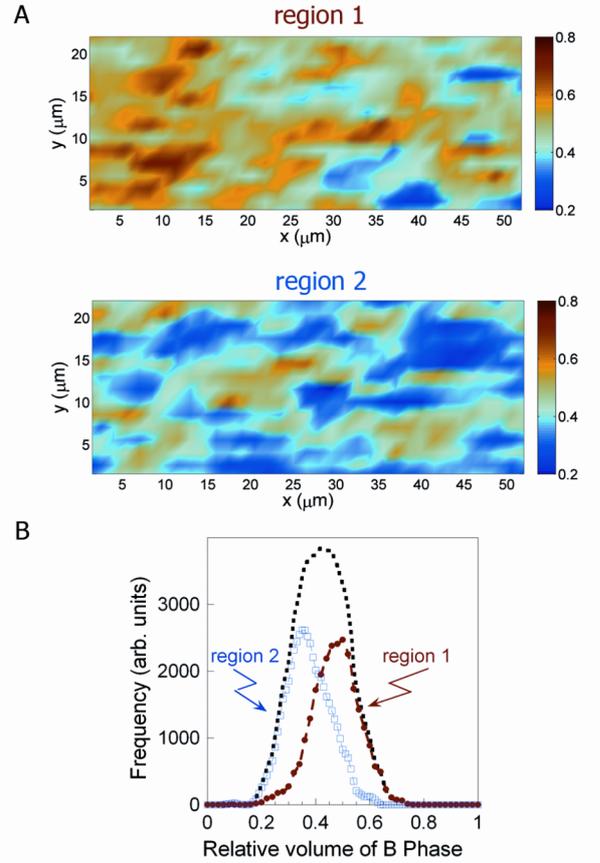

**Figure 4:** (Color online) (A) shows the spatial distribution of ratio of the compressed and the expanded phase in two different regions of size $22 \times 55 \mu m^2$ of the $K_{0.8}Fe_{1.6}Se_2$ crystal. The crystallographic a and b axis are along the horizontal and the vertical directions. The intensity of the compressed phase $I(c)$ and of the expanded phase $I(e)$ is integrated over square subareas of the image recorded by the CCD detector in reciprocal lattice (r.l.u) and their ratio $[I(c) - I_0]/[(I(e) - I_0) + (I(c) - I_0)]$ is plotted at each (x,y) point of the sample surface. The intense orange red (lighter gray) peaks in the two-dimensional color map represent locations of the sample with a dominant compressed phase, and black (blue-color online) indicates spots of dominant expand phase. (B) shows the frequency (arbitrary units) of the ratio values between the compressed and the expanded phase in different surface regions (relative volume of the collapsed phase) is plotted for the first (blue filled circles) and second (red empty square) surface regions, representing the distribution in different regions.

Fig. 4 shows the spatial mapping of the ratio of the two phases, revealing changes from spot to spot. There are regions where the B phase is dominant and other large regions where the A phase is dominant. Fig. 4 shows two such regions of the $K_{0.8}Fe_{1.6}Se_2$ crystal of size $22 \times 55$ $\mu m^2$ mapped using scanning x-ray nanofocus diffraction in reflection mode. The intense red-yellow (gray-lighter gray) peaks in the two-dimensional color maps (Fig. 4) represent locations with predominance of the B phase. In particular, the two maps show an over all weights for the B-A phase to be of 50%-50% and 30%-70% respectively. This value appears to be consistent with the values estimated using XRD in transmission mode, probing a much larger bulk volume (with values to be almost 30%-70 %).

The present observation of an intrinsic phase separation with the coexistence of an expanded and a compressed phase suggests the importance of inhomogeneity in the $A_xFe_{2-y}Se_2$ systems. To have an estimate of the coherence length associated with the domains of the two phases, we have used deconvolution of the peak profiles (Figure 3 lower panel). The extracted domain size from the peak profile analysis, measured by the diffraction coherence length, is found to be $\xi^l = 6 \pm 1 nm$. This small size of the homogenous phase points towards a frustrated nanoscale phase separation. Moving the nano-x-ray beam on the sample surface we find that the ratio between the intensity of the reflections probing the ratio between the compressed B phase and the A phase changes in a dramatic way. As can be seen in Fig. 3, the relative probability of the compressed versus expanded phase is ~30% in one region, ~50% in other, and is ~70% in another region, underlining the frustrated nature of the phase separation. It should be recalled that the microscopic observation of phase separation has been provided by an earlier transmission electron study on this system [13]. The experimental tool used for the present work has allowed us to study the system in different length scales, thus enabling a systematic investigation on the nature of phase separation.

In conclusion, we have reported the imaging of the complex phase separation in the recently discovered $K_xFe_{2-y}Se_2$ superconducting system by mapping the lattice order using nanofocus scanning x-ray diffraction. An expanded phase with superlattice modulation is found to coexist with a compressed phase, in a multiscale frustrated spatial distribution, having direct consequence on the nonmagnetic and magnetic properties of the system. The observed complex landscape of the phase separation recalls the case of a superoxygenated $La_2CuO_4$ [19,32] in which a fractal distribution of dopants appear to enhance the superconductivity. Indeed, the phase separation shown in $K_xFe_{2-y}Se_2$ chalcogenides appears to be an important feature for the understanding of the physics of superconductivity in these materials, similar to the one discussed for the cuprates and other related systems [18,22-31].

**Acknowledgements.** We thank the ESRF XRD 13 beamline staff in Grenoble and the Elettra XRD beamline staff in Trieste for experimental help.